\newif\ifeprint\eprinttrue
\documentclass
[rmp,\ifeprint twocolumn\else preprint\fi
,amsfonts,amssymb,amsmath,showkeys,letterpaper,raggedbottom]{revtex4}
\date{{\year=2005\month=6\day=23\today}; revised
{\year=2005\month=8\day=26\today} and \today}

\def\TITLE{Quaternions in molecular modeling}
\def\KEYWORDS{quaternions, rotations, molecular simulation}

\usepackage{times}
\usepackage[bookmarks=false]{hyperref}
\hypersetup{pdftitle={\TITLE},
pdfauthor={C. F. F. Karney},
pdfkeywords={\KEYWORDS}}

\bibpunct{[}{]}{;}{n}{}{,}

\makeatletter\def\raggedcolumn@skip{\vskip\z@\@plus.0001fil\relax}\makeatother

\def\hrefx#1#2#3{\href{#1}{#2}\penalty0\href{#1}{#3}}

\DeclareMathAlphabet{\mathbsf}{OT1}{cmss}{bx}{n}
\def\v#1{\mathbf{#1}}
\def\q#1{\mathbsf{#1}}
\def\V#1{\mathbf{#1}}
\def\Q#1{\mathbsf{#1}}
\def\abs#1{\left|#1\right|}
\def\absb#1{\bigl|#1\bigr|}
\def\ave#1{\langle#1\rangle}
\def\mean#1{\langle\!\langle#1\rangle\!\rangle}
\def\conj#1{\overline{#1}}
\def\cross#1{\V I \times #1}
\def\transp{^{\mathsf T}}

\advance\textheight 6pt \advance\topmargin -3pt

\begin{document}

\ifeprint
\noindent\mbox{\begin{minipage}[b]{\textwidth}
\begin{flushright}\tt\footnotesize
Link: \href{http://charles.karney.info/biblio/quat.html}
           {http://charles.karney.info/biblio/quat.html}\par\vspace{0.5ex}
E-print: \href{http://arxiv.org/abs/physics/0506177}
                             {arXiv:physics/0506177}\par
\vspace{2ex}
\end{flushright}
\end{minipage}
\hspace{-\textwidth}}
\fi

\title{\TITLE}
\ifeprint
\author{\href{http://charles.karney.info}{Charles F. F. Karney}}
\else
\author{Charles F. F. Karney}
\fi
\email{ckarney@sarnoff.com}
\affiliation{\href{http://www.sarnoff.com}{Sarnoff Corporation},
  Princeton, NJ 08543-5300}

\begin{abstract}

Quaternions are an important tool to describe the orientation of a
molecule.  This paper considers the use of quaternions in matching two
conformations of a molecule, in interpolating rotations, in performing
statistics on orientational data, in the random sampling of rotations,
and in establishing grids in orientation space.  These examples show
that many of the rotational problems that arise in molecular modeling
may be handled simply and efficiently using quaternions.

\end{abstract}

\maketitle

\section{Introduction}

Quaternions were introduced in the mid-nineteenth century by Hamilton
\nocite{hamilton44}\cite{hamilton44,hamilton47} as an extension of
complex numbers and as a tool for manipulating 3-dimensional vectors.
Indeed Maxwell used them to introduce vectors in his exposition of
electromagnetic theory \cite[\S\S10--11]{maxwell92}.  However, unlike
complex numbers which occupy a central role in the development of
algebra, quaternions found no similar place in mathematics and, with the
introduction of modern vector notation by Gibbs \cite{gibbs01},
quaternions fell out of favor by the end of the nineteenth century.
Nevertheless, quaternions excel as a way of representing rotations of
objects in 3-dimensional space.  They are economical to work with (both
in terms of storage and computation); but more importantly they offer a
clean conceptual framework which allow several problems involving
rotations to be easily solved.

Basic quaternion algebra is well covered in Hamilton's papers
\cite{hamilton44,hamilton47}, which are both accessible and readable.
These papers may be supplemented with a wealth of on-line resources
\nocite{wikipedia}\cite{wikipedia,mathworld}.  Many authors over the
past 20 years have ``rediscovered'' the application of quaternions to
rotations and it is with some trepidation that this author inflicts
another paper on the subject on the scientific community.  However,
within the molecular modeling community, quaternions are quite narrowly
applied.  This paper therefore briefly reviews quaternion algebra and
then describes their applications to a broad range of rotational
problems in molecular modeling.  Much of this material has appeared
before---but often scattered about in journals for fields unrelated to
molecular modeling.  I have, therefore, endeavored to organize the
material, to generalize it, and to present it with a consistent
notation, with the hope this affords a deeper appreciation of the power
of quaternions in describing rotations and encourages their wider
adoption in molecular modeling.

The outline of this paper is as follows.  After introducing quaternions
and their use in describing rotations, we tackle various applications.
First we review the quaternion method for computing the least-squares
fit of two conformations of the same molecule.  We also see how to
include molecular inversions and discuss why the least-squares fit is a
poor choice to describe the orientation of a flexible molecule.  We next
show how to interpolate smoothly between two orientations and that this
corresponds to rotating the molecule at constant angular velocity.  In
order to carry out statistics on orientational data, we give a robust
definition of the mean orientation showing how to transform the
deviations from the mean to 3-dimensional space so that familiar
statistical tools may be employed.  In Monte Carlo applications, we need
to be able to select a random orientation uniformly; we show that this
is trivially accomplished in quaternion space and we also consider the
problem of making random incremental rotations.  Finally, it is
frequently useful to impose a grid on orientation space and we
illustrate how this may be done with applications to quadrature and
searching.

\section{Quaternions}

The original notation for quaternions \cite{hamilton44} paralleled the
convention for complex numbers
\[
\q q = q_0 \q u + q_1 \q i +  q_2 \q j +  q_3 \q k,
\]
which obey the conventional algebraic rule for addition and
multiplication by scalars (real numbers) and which obey an associative
non-commutative rule for multiplication where $\q u$ is the identity
element and
\begin{eqnarray*}
&\q i^2 = \q j^2 = \q k^2 = - \q u, \\
&\q i\q j = - \q j\q i = \q k,\quad
\q j\q k = - \q k\q j = \q i, \quad
\q k\q i = - \q i\q k = \q j.
\end{eqnarray*}
It is frequently useful to regard quaternions as an ordered set of 4
real quantities which we write as
\begin{equation}\label{qord}
\q q = [q_0, q_1, q_2, q_3],
\end{equation}
or as a combination of a scalar and a vector
\begin{equation}\label{qsv}
\q q = [q_0, \v q],
\end{equation}
where $\v q = [q_1, q_2, q_3]$.  A ``scalar'' quaternion has zero
vector part and we shall write this as $[q_0, \v 0] = q_0 \q u = q_0$.
A ``pure'' quaternion has zero scalar part $[0,\v q]$.

In the scalar-vector representation, multiplication becomes
\[
\q p \q q = [p_0 q_0 - \v p\cdot\v q,\;
             p_0\v q + q_0\v p + \v p \times \v q],
\]
where ``$\cdot$'' and ``$\times$'' are the vector dot and cross products.
The conjugate of a quaternion is given by
\[
\conj{\q q} = [q_0, -\v q];
\]
the squared norm of a quaternion is
\[
\abs{\q q}^2 = \q q\conj{\q q} = q_0^2 + q_1^2 + q_2^2 + q_3^2,
\]
and its inverse is
\[
\q q^{-1} = \conj{\q q} / \abs{\q q}^2.
\]
Quaternions with $\abs{\q q} = 1$ are called unit quaternions, for
which we have $\q q^{-1} = \conj{\q q}$.

The quaternion $\q q$ can also be represented as a $2\times2$ complex
matrix,
\[
\left(
\begin{array}{cc}
 q_0 + i q_1& q_2 + i q_3\\
-q_2 + i q_3& q_0 - i q_1
\end{array}
\right),
\]
or as a $4\times4$ real matrix,
\begin{equation}\label{rightmult}
\Q Q(\q q) = \left(
\begin{array}{cccc}
 q_0& q_1& q_2& q_3\\
-q_1& q_0&-q_3& q_2\\
-q_2& q_3& q_0&-q_1\\
-q_3&-q_2& q_1& q_0
\end{array}
\right);
\end{equation}
in these forms, quaternion multiplication becomes matrix multiplication.

The notation we adopt here is to use light-face italics for scalar
quantities, bold roman for 3-dimensional vectors and $3\times3$
matrices, bold sans serif for quaternions and $4\times4$ matrices.
Quaternion multiplication is indicated by $\q p \q q$, while ``$\cdot$''
is used to indicate matrix-vector and vector-vector (including
quaternion-quaternion) contractions and in this context $\q q$ and $\v
v$ are treated as {\em column} vectors.  Thus, we may write $\abs{\q
q}^2 = \q q\transp\cdot \q q$.  We also find that $\q p \q q = \q
p\transp \cdot \Q Q(\q q)$, with $\Q Q$ given by eq.~(\ref{rightmult}).
Consistent with eqs.~(\ref{qord}) and (\ref{qsv}), we shall number
quaternion indices starting at 0 and vector indices from 1.

\section{Rotations}

The chief application of quaternions to molecular modeling lies in their
use to represent rotations.  Consider a unit quaternion
\begin{equation}\label{rot0}
\q q = [\cos(\theta/2), \v v \sin(\theta/2)],
\end{equation}
where $\abs{ \v v} = 1$, and define an operator $R_{\q q}$ on
3-dimensional vectors by
\begin{equation}\label{rot}
[0,R_{\q q}(\v x)] = \q q\,[0, \v x]\,\conj{\q q}.
\end{equation}
Multiplying out the quaternion product, we find
\[
R_{\q q}(\v x) = \V R(\q q)\cdot \v x,
\]
where $\V R(\q q)$ is the tensor
\begin{eqnarray}
\V R(\q q) &=& (q_0^2 - \abs{\v q}^2) \V I
                + 2 \v q \v q + 2 q_0 \cross{\v q}\label{mat1}\\
&=& \v v\v v
  + \cos\theta(\V I - \v v \v v)
  + \sin\theta \, \cross{\v v},\label{mat2}
\end{eqnarray}
where $\v a \v a$ is the parallel projector [$(\v a \v a) \cdot \v b =
(\v a \cdot \v b) \v a$] and
$\cross{\v a}$ is the cross operator [$(\cross{\v a}) \cdot \v b =
\v a\times \v b$] \cite[\S113]{gibbs01}.  Equation (\ref{mat2}) is the
conventional tensor representation for a right-handed rotation of
$\theta$ about an axis $\v v$ through the origin \cite[\S126]{gibbs01}.
Equation (\ref{mat1}) may be written in component form as
\begin{eqnarray}
\ifeprint\hspace{-1.5em}\fi
\V R(\q q) &=&
\ifeprint\nonumber\\ &&\hspace{-3em} \fi
\left(
\begin{array}{ccc}
1-2q_2^2-2q_3^2 & 2q_1q_2-2q_0q_3 & 2q_1q_3+2q_0q_2 \\
2q_2q_1+2q_0q_3 & 1-2q_3^2-2q_1^2 & 2q_2q_3-2q_0q_1 \\
2q_3q_1-2q_0q_2 & 2q_3q_2+2q_0q_1 & 1-2q_1^2-2q_2^2 \\
\end{array}
\right).\label{mat3}
\end{eqnarray}

The definition, eq.~(\ref{rot}), gives $R_{\q p}(R_{\q q}(\v x)) = R_{\q
p\q q}(\v x)$, so that $\q p\q q$ corresponds to composing rotations
(with the rotation by $\q q$ performed first).  We also find that $R_{\q
q} = R_{-\q q}$; i.e., $\q q$ and $-\q q$ give the same
rotation---changing the sign of $\q q$ is equivalent to increasing
$\theta$ by $2\pi$ in eq.~(\ref{rot0}).  Unit quaternions satisfy $q_0^2
+ q_1^2 + q_2^2 + q_3^2 = 1$ and the quaternion representation of
rotations are as points on a hypersphere $\mathbb S^3$ with opposite
points identified.  For future reference, we note that the
(three-dimensional) area of
$\mathbb S^3$ is $2\pi^2$.

Because $\q q$ and $-\q q$ give the same rotation, some care needs to be
taken when comparing two orientations represented by $\q q_a$ and $\q
q_b$.  The rotation, $\q q = \q q_b \conj{\q q_a}$, moves from $\q q_a$
to $\q q_b$.  When inverting eq.~(\ref{rot0}) to determine the rotation
angle $\theta$ between the two orientations, we should, if necessary,
change the sign of $\q q$ to ensure that $q_0 \ge 0$, so that $\theta
\in [0,\pi]$.  A simple metric for closeness is given by $\cos(\theta/2)
= \abs{\q q_a\transp \cdot \q q_b}$.

Describing rotations with quaternions has a number of benefits.  They
offer a compact representation of rotations.  Compared to Euler angles,
they are free of singularities.  Rotations may be composed more
efficiently using quaternions than by matrix multiplication.  Also in
contrast to rotation matrices, it is easy to maintain a quaternion's
unit normalization (merely divide it by $\abs{ \q q}$).  However the
chief benefit is that the representation of a rotation as point on
$\mathbb S^3$ allows us to derive many important results concerning
rotations in a simple coordinate-free way.

There is one application where the matrix representation of rotations is
more efficient that the quaternion representation.  If we wish to apply
the same rotation to many points, then we should form the rotation
matrix using eq.~(\ref{mat1}) and transform the points by matrix
multiplication.

The conventional representation for rotations that is most closely
allied to quaternions is the axis-angle representation, where the
rotation is given by a vector $\v s = \theta\v v$ which denotes a
rotation of $\theta = \abs{\v s}$ about an axis $\v v = \v s/\abs{\v
s}$.  It is useful to have an analytic relation between the quaternion
and axis-angle representations and this is provided by the quaternion
exponential \cite{hamilton44},
\begin{equation}\label{expdef}
\exp([0, \v s/2]) = \q q,
\end{equation}
where $\q q$ is given by eq.~(\ref{rot0}), This definition of the
exponential follows from its series expansion.  Similarly the inverse
operation is given by the quaternion logarithm
\begin{equation}\label{logdef}
\ln\q q = [0, \v s/2 + 2\pi n \v v],
\end{equation}
where $n$ is an integer.

It is useful here to make a distinction between ``orientation'' and
``rotation''.  We imagine that our molecule has some arbitrary but
definite reference state.  We apply a rotation and a translation
(jointly referred to as a ``displacement'') to this reference state and
so bring the molecule to a new orientation and position (jointly
referred to as a ``configuration'').

\section{Least-squares fit}

Given two conformations of the same molecule, it is often useful to be
able to determine how close the conformations are.  In order to do this,
we can rigidly move one conformation so that it nearly coincides up with
the other and then determine the difference in the positions of the
corresponding atoms.  Thus, if we are given two sets of atomic positions
$\{\v x_k\}$ and $\{\v y _k\}$ with $k \in [1,N]$ together with a set of
atomic ``weights'' $\{w_k\}$, we wish to determine the (rigid)
displacement $T$ which minimizes
\begin{equation}\label{Emin}
E = \frac1W \sum_k w_k \abs{ \v y_k - T(\v x_k) }^2,
\end{equation}
where $W = \sum_k w_k$.  Here $w_k$ is merely a statistical weight of an
atom---it is not necessarily related to the atomic mass.  The two sets
of atomic positions are ordered which presumes that we can identify
corresponding atoms.  (This is not necessarily a simple matter, if, for
example, we are dealing with a molecule with several identical
branches.)  The displacement $T = (\q q, \v d)$ may be expressed as a
rotation about an axis through the origin followed by a translation,
i.e., $T(\v x) = R_{\q q}(\v x) + \v d$.

This problem has been considered by many authors and a review of various
approaches is given by Flower \cite{flower99}.  Using quaternions to
describe the rotation leads to an elegant and robust solution.  An early
use of quaternions in this context is to solve the problem formulated by
Wahba \nocite{wahba65}\cite{wahba65,wahba66}, the determination of the
attitude of a spacecraft given the directions of several objects
relative to the craft.  The resulting ``$\q q$-method'' is described by
Keat \cite[\S A.3]{keat77} and by Lerner \cite[\S12.2.3]{lerner78} who
both credit the invention of the method to Paul B. Davenport (1968).
The generalization to matching points (as opposed to directions) was
considered by Faugeras and Hebert \cite{faugeras83} who independently
found the same method for determining the orientation.  Their method was
subsequently rediscovered by Horn \cite{horn87}, by Diamond
\cite{diamond88}, and by Kearsley \cite{kearsley89}.  The derivation of
Faugeras and Hebert is one of the clearest, and we briefly summarize it
here including the straightforward generalization of including arbitrary
weights $w_k$.

If we demand that the variation of $E$ with respect to $\v d$ vanish, we
find that
\begin{equation}\label{translation}
\v d = \ave{\v y} - R_{\q q}(\ave{\v x}),
\end{equation}
where $\ave{\ldots}$ denotes the sample average,
\begin{equation}\label{average}
\ave X = \frac1W \sum_k w_k X_k.
\end{equation}
Equation (\ref{Emin}) may now be written as
\begin{equation}\label{Emin1}
E = \frac1W \sum_k w_k \abs{ \v y'_k - R_{\q q}(\v x'_k) }^2,
\end{equation}
where $\v x'_k = \v x_k - \ave{\v x}$ and $\v y'_k = \v y_k - \ave{\v
y}$.  Using eq.~(\ref{rot}), eq.~(\ref{Emin1}), becomes
\begin{equation}\label{Emin2}
E = \frac1W \sum_k w_k \absb{ [0, \v y'_k] -
  \q q\,[0, \v x'_k]\,\conj{\q q} }^2.
\end{equation}
Because, the norm of a quaternion is unchanged on multiplying it by a
unit quaternion, we may right-multiply the kernel of eq.~(\ref{Emin2})
by $\q q$ to give
\begin{equation}\label{Emin3}
E = \frac1W \sum_k w_k
 \absb{ [0, \v y'_k]\,\q q -  \q q\,[0, \v x'_k] }^2.
\end{equation}
We need to minimize eq.~(\ref{Emin3}) subject to the constraint $\abs{\q
q} = 1$.  Because the kernel is linear in $\q q$, it can be written as
\begin{equation}
 [0, \v y'_k]\,\q q -  \q q\,[0, \v x'_k] = \Q A_k \cdot \q q,
\end{equation}
where $\Q A_k$ is a $4\times 4$ skew matrix
\[
\Q A_k = \Q A(\v y'_k + \v x'_k, \v y'_k - \v x'_k),
\]
with
\begin{eqnarray*}
\Q A(\v a, \v b) &=&\left(
\begin{array}{cc}
0 & -\v b\transp\\
\v b & \cross{\v a}
\end{array}
\right)\\
&=& \left(
\begin{array}{cccc}
 0  &-b_1&-b_2&-b_3\\
 b_1& 0  &-a_3& a_2\\
 b_2& a_3& 0  &-a_1\\
 b_3&-a_2& a_1& 0
\end{array}
\right).
\end{eqnarray*}
Substituting this into eq.~(\ref{Emin3}), we obtain
\begin{equation}\label{Emin4}
E = \frac1W \sum_k w_k\,
    \q q\transp \cdot \Q A_k\transp \cdot \Q A_k \cdot \q q
  = \q q\transp \cdot \Q B \cdot \q q,
\end{equation}
where $\Q B = \ave{\Q A_k\transp \cdot \Q A_k}$ is a $4\times 4$
symmetric matrix which has real eigenvalues, $0 \le \lambda_0 \le
\lambda_1 \le \lambda_2 \le \lambda_3$.  Setting $\q q$ to the
eigenvector corresponding to $\lambda_0$ gives the minimum value for $E
= \lambda_0$.

In summary, the best fit is achieved by subtracting the mean positions
from the original sets of points to give $\{\v x'_k\}$ and $\{\v
y'_k\}$, forming the matrices $\Q A_k$ and $\Q B$, and determining the
minimum eigenvalue $\lambda_0$ of $\Q B$.  The optimal rotation is given
by setting $\q q$ to the corresponding eigenvector of $\Q B$ and the
optimal translation is found from eq.~(\ref{translation}).  The mean
squared error for this fit is $\lambda_0$.

This procedure has two attractive features.  The rotation obtained is a
proper rotation (without an inversion); this is usually the desired
result.  Secondly, degenerate molecules are treated satisfactorily.  For
example if one or both of the sets $\{\v x_k\}$ and $\{\v y _k\}$ is
collinear, then the best fit is no longer unique.  The result will be
that there will be multiple minimum eigenvalues of $\Q B$ with distinct
eigenvectors.  The general solution is obtained by setting $\q q$ to a
linear combination of these eigenvectors.  The method does require
finding the eigenvalues and eigenvectors of a $4\times 4$ matrix.
However there are many numerical libraries
\nocite{lapack}\nocite{gsl}\cite{lapack,gsl,pozo97} which solve such
problems and the results are accurate to round-off for small symmetric
matrices such as $\Q B$.  A fast method of determining just the required
eigenvector and associated eigenvalue in order to determine the attitude
of a spacecraft is given in \cite[\S III]{shuster81}.  However, in
applications to molecular modeling, it is probably preferable merely to
invoke a library eigenvector routine.

Horn \cite{horn87} considered including a scaling in the transformation
$T$ in eq.~(\ref{Emin}).  This is quite easily accommodated.  However
there seems little need to include such an effect in molecular modeling.

Diamond \cite{diamond90} considers the case where inversions are
allowed.  This is easily achieved by substituting $-\v x'_k$ for $\v
x'_k$ in eq.~(\ref{Emin1}).  Equation~(\ref{Emin4}) then involves a
matrix $\Q B'$ where
\begin{equation}
\Q B' =
2 \ave{ \abs{\v x'}^2 + \abs{\v y'}^2 } \Q I
 - \Q B.
\end{equation}
Consequently the rotation giving the best inverted fit is the
eigenvector with the {\em greatest} eigenvalue of $\Q B$, $\lambda_3$.
Because the sum of the eigenvalues of $\Q B$ is its trace, $4
\ave{\abs{\v x'}^2 + \abs{\v y'}^2}$, we can express the mean squared
error for the inverted fit as $\frac12(\lambda_0 + \lambda_1 + \lambda_2
- \lambda_3)$.  Thus, once the eigenvalues of $\Q B$ have been computed
we immediately determine whether the inverted fit will be better than
the proper fit.

Coutsias {\em et al.}\ \cite{coutsias04} provide an interesting
extension of this method.  Suppose the atomic positions $\{\v x_k\}$
represent a model of a molecule which depends on a set of parameters
$\{\alpha_i\}$, for example, the torsion angles of a protein backbone.
By considering the gradient of $E$ in parameter space $\partial
E/\partial \alpha_i$, they provide a method for determining the
parameter values which result in the best fit to a given crystal
structure.

One other interesting consequence of the result for the best fit is
that the rotation is not a continuous function of the configurations of
the molecules.  Let us suppose that $\{\v x_k\}$ gives the position of
the atoms in a molecule in some predefined configuration and suppose
that $\{\v y _k\}$ gives the atom positions during the course of a
dynamical simulation of the molecule.  If the forces acting on the atoms
are finite then $\v y_k$ is a $C^1$ function (twice differentiable).
During the course of the deformation of the molecule, $\Q B$ and its
eigenvalues change.  In the typical case, the two smallest eigenvalues
exchange roles and $\q q$ switches from one direction in $\mathbb R^4$
to an orthogonal direction.  This results in the orientation of the best
fit changing discontinuously by $180^\circ$.

In modeling a flexible molecule, it is frequently useful to separate the
external degrees of freedom, namely position and orientation, from the
internal degrees of freedom.  This allows, for example, translational
and rotational symmetry to the system to be enforced and correlations
between the motions of atoms within a molecule to be studied.  This begs
the question of how best to define the position and orientation of a
molecule.  Taking the position to be the center of mass is often the
obvious choice.  The position (so defined) evolves according to Newton's
second law driven by the total force on the molecule.  It is not
possible to keep track of the orientation in an analogous fashion by
integrating the total angular momentum, because flexible bodies can
change their orientation with zero angular momentum---witness the
ability of a cat always to land on its feet.  A possible definition of
the orientation is the best fit orientation to a reference conformation;
i.e., we define $\q o_R(A)$ as the best fit orientation, expressed as a
quaternion, of the molecule in conformation $A$ relative to a reference
conformation $R$.  Here again this choice has the attractive feature
that the whole molecule is included in the definition.  There are two
problems with this prescription.  Firstly, the difference in
orientations between two conformations $A$ and $B$ depends, in general,
on the choice of reference conformation, namely
\[
\q o_R(B) \conj{\q o_R(A)} \ne \q o_S(B) \conj{\q o_S(A)}.
\]
(This is easily demonstrated for simple triatomic molecules.)  Thus this
definition of orientation entails a degree of ``arbitrariness'' absent
in our definition of position.  A second more serious defect arises from
the discussion in the previous paragraph.  Recovering the actual
configuration of the molecule from the orientation defined in this way
is numerically unstable (by a flip of $180^\circ$!)\ whenever the lowest
eigenvalues cross.  This would also lead to large and discontinuous
apparent internal motions of the molecule with small changes in the
atoms' true positions.  A better choice would therefore be to make the
fit to some rigid (or nearly rigid) subcomponent of the molecule
\cite{hunenberger95}.  Although this still yields an arbitrary
definition of orientation (depending on the choice of reference
subcomponent), the resulting orientation varies continuously under
continuous deformations of the molecule.  An extensive discussion of how
to separate the orientation from the internal motions of a flexible
molecule is given by Littlejohn and Reinsch \cite{littlejohn97}.

\section{Interpolating rotations}

The power of the quaternion representation of rotations is evident when
we consider the problem of interpolating between two orientations of a
molecule.  (This application might arise in the animation of a molecular
simulation.)  Suppose we wish to interpolate between $\q q_a$ and $\q
q_b$.  Because these quaternions and their interpolants lie on the unit
sphere $\mathbb S^3$, the shortest path will be a great circle whose
parametric equation is given by \cite{shoemake85}
\begin{equation}\label{slerp}
\q q(\phi) =
\frac{ \q q_a \sin(\theta - \phi) + \q q_b \sin(\phi) }{\sin(\theta)},
\end{equation}
where $\cos\theta = \q q_a\transp\cdot \q q_b$.  In the computer
animation community this ``spherical linear interpolation'' operation is
denoted by $\mathop{\mathrm{Slerp}}(\q q_a, \q q_b; u) = \q q(u \theta)$
\cite{shoemake85}.  As $\phi$ is increased from $0$ to $2\pi$, $\q
q(\phi)$ becomes successively $\q q_a$, $\q q_b$, $-\q q_a$, $-\q q_b$,
and finally returns to $\q q_a$.  During this operation the
corresponding 3-dimensional rotation has increased by $4\pi$.  If $\q
q_a\transp\cdot \q q_b \ge 0$, then $0\le \phi \le \theta$ takes $\q
q(\phi)$ smoothly from $\q q_a$ to $\q q_b$.  If, on the other hand, $\q
q_a\transp\cdot \q q_b < 0$, then a shorter path is found with $0\ge
\phi \ge \theta - \pi$ which takes $\q q(\phi)$ smoothly from $\q q_a$
to $-\q q_b$.

Equation (\ref{slerp}) is derived using simple geometrical arguments
applied to $\mathbb S^3$ and the same result is obtained for the
great-circle interpolation for $\mathbb S^n$.  For $\mathbb S^3$, the
result can also be expressed as
\[
\q q(\phi) = (\q q_b \conj{\q q_a})^{\phi/\theta} \q q_a.
\]
This relation has the interpretation: rotate to $\q q_a$ and then rotate
a fraction $\phi/\theta$ to the path from $\q q_a$ to $\q q_b$.  The
operation $\q q^u$ is defined by  \cite{hamilton44}
\[
\q q^u = \exp(u \ln \q q).
\]

In fact this interpolation scheme results in the molecule undergoing
rotation at constant angular velocity.  In order to show this, consider
a body rotating at $\omega$ about a unit axis $\v v$.  The evolution of
the orientation $\q q$ satisfies the differential equation
\begin{equation}\label{ode}
\dot{\q q} = [0, (\omega/2) \v v]\,\q q.
\end{equation}
This is easily solved (e.g., by using finite differences and passing to
the limit $\delta t \rightarrow 0$) to give
\begin{eqnarray*}
\q q(t) &=& \exp([0, (\omega t/2) \v v])\q q(0)\\
&=&[\cos(\omega t/2), \v v \sin(\omega t/2)]\, \q q(0),
\end{eqnarray*}
which agrees with eq.~(\ref{slerp}) with the substitutions $\phi =
\omega t/2$, $\q q_a = \q q(0)$ and $\q q_b = [0, \v v]\,\q q(0)$.

If we wish to interpolate between two configurations of a rigid
molecule, we are free to specify a point, $\v x_0$, in the reference
molecule which will move with constant velocity.  If the initial and
final configurations are given by $T_a = (\q q_a, \v d_a)$ and $T_b =
(\q q_b, \v d_b)$, with $\q q_a\transp\cdot \q q_b \ge 0$, then the
required interpolation is achieved by increasing $u$ from $0$ to $1$
with the orientation given by $\q q(u\theta)$ and the translation given
by
\[
(\v d_a + R_{\q q_a}(\v x_0))(1-u)
+ (\v d_b + R_{\q q_b}(\v x_0))u
-R_{\q q(u\theta)}(\v x_0).
\]

\section{Mean Orientation}

The mean of directional quantities has frequently presented difficulties
\cite{mardia99}.  Let us assume we have $N$ samples of some directional
quantity with weights $w_k$ for $k \in [1,N]$ and $\sum_k w_k = W$.  In
the case where the samples are angles (e.g., the dihedral angles of a
molecular bond) or directions (e.g., the orientations of a diatomic
molecule), there is a well established procedure
\cite[\S2.2.1,~\S9.2.1]{mardia99}: express the directions as unit
vectors in $\mathbb R^2$ or $\mathbb R^3$, $\v n_k$, and determine
$\ave{\v n}$ where we take the sample average according to
eq.~(\ref{average}).  Now the mean direction is given by $\mean{\v n} =
\ave{\v n} / \abs{\ave{\v n}}$, while $1-\abs{\ave{\v n}}$, a quantity
lying in $[0,1]$, is the ``circular variance'' \cite[\S2.3.1]{mardia99}
or ``spherical variance'' \cite[\S9.2.1]{mardia99}.  Here $\ave{\ldots}$
is defined as a simple weighted arithmetical average,
eq.~(\ref{average}), while $\mean{\ldots}$ denotes the physically
relevant mean of a quantity.

This procedure cannot be directly applied to unit quaternions used to
represent rotations because of the indistinguishability of $\pm\q q$.
Instead, we view $\{\q q_k\}$ as {\em axes}
\cite[\S1.1,~\S9.1]{mardia99} in $\mathbb R^4$, and define $\mean{\q q}$
as the unit quaternion about which the weighted moment of inertia of
$\{\q q_k\}$ is minimum \cite[\S3]{rancourt00}.  Thus we wish to
minimize
\begin{eqnarray*}
L & = & \frac1W \sum_k w_k\,
  \q q_k\transp \cdot (\Q I - \mean{\q q} \mean{\q q}\transp)
     \cdot \q q_k \\
  & = & \frac1W \sum_k w_k\,
  \mean{\q q}\transp \cdot (\Q I - \q q_k \q q_k\transp)
     \cdot \mean{\q q} \\
  & = & \mean{\q q}\transp \cdot (\Q I - \ave{\q q \q q\transp})
     \cdot \mean{\q q}.
\end{eqnarray*}
The minimum value of $L$ is given by the minimum eigenvalue of $\Q I -
\ave{\q q \q q\transp}$ and $\mean{\q q}$ is corresponding eigenvector.
The resulting $L$, which is a quantity lying in $[0, \frac34]$, then
provides a measure of the variance of the rotations.  This definition of
the mean has a number of desirable properties: it is invariant when the
signs of the $\q q_k$ are changed; it is independent of the order of the
samples; and it transforms properly if the samples are transformed.

This prescription can also be applied to determine the mean direction of
objects whose symmetry makes $\v n$ and $-\v n$ indistinguishable (for
example, the orientation of the diatomic molecule $\mathrm N_2$).

Suppose we wish to determine the mean configuration of a rigid molecule,
i.e., the mean of $\{T_k = (\q q_k, \v d_k)\}$.  We are free to choose a
point $\v x_0$ in the reference molecule whose position in the mean
configuration coincides with its mean position.  (Compare this with the
discussion of interpolating configurations in the previous section.)  A
suitable definition for the mean configuration is then
\begin{equation}\label{meanconfig}
\mean T = (\mean{\q q},
   \ave{\v d} + \ave{R_{\q q}(\v x_0)} - R_{\mean{\q q}}(\v x_0)).
\end{equation}

Frequently, we need more precise information about the distribution of
configurations than its variance.  We might need to know how much the
rotation about different axes are correlated or whether rotational and
translational motions are coupled.  It is also desirable to be able to
fit model distributions to a set of samples.  For these purposes, it is
useful to be able to map rotations onto $\mathbb R^3$ so that standard
statistical tools can be employed.  We require that the mapping be
measure preserving (constant Jacobian) to simplify the use of the
transformed rotations.

We have already introduced the axis-angle representation of rotations.
We may make the restriction $\abs{\v s} \le \pi$ and so map the
hemisphere $q_0\ge0$ of $\mathbb S^3$ onto a ball of radius $\pi$ in
$\mathbb R^3$.  Unfortunately, the mapping, eq.~(\ref{expdef}), does not
have constant Jacobian.  We can correct this by defining \cite{karney05a}
a new ``turn''
vector $\v u$ with the properties
\begin{subequations}
\label{lambert}
\begin{eqnarray}
\v u & \parallel & \v s, \\
\abs{\v u} &=& \biggl(\frac{\abs{\v s} - \sin\abs{\v s}}\pi\biggr)^{1/3}.
\end{eqnarray}
\end{subequations}
This is an extension of the Lambert azimuthal equal-area projection
providing a measure-preserving mapping of the hemisphere $q_0\ge0$ of
$\mathbb S^3$ onto the unit ball $\mathbb B^3$.  Equation
(\ref{lambert}) is well behaved at the boundary, $\abs{\v u} = 1$;
however on this boundary antipodal points are identified.  The inverse
mapping has an infinite derivate at $\abs{\v u} = \sqrt[3]{2n}$ for
integer $n \ne 0$ which corresponds to shells in $\v u$ space which map
to the origin.  This inverse of eq.~(\ref{lambert}) is easily
implemented via Newton's method supplemented by a Taylor series at the
origin and at $\sqrt[3]{2n}$.

This mapping was introduced \cite{karney05a} to allow distributions of
orientations to be fit using a mixture of Gaussians \cite{dempster77}.
Given a set of sample orientations $\{\q q_k\}$, we compute the mean
orientation, $\mean{\q q}$.  The deviations of the samples from the mean
are then given by the rotations $\{\q q_k \conj{\mean{\q q}}\}$ and
these are mapped to a set of turns $\{\v u_k\}$.  Because these are
points in $\mathbb R^3$, we may fit them with a 3-dimensional Gaussian
with zero mean and with covariance matrix $\ave{\v u\transp\v u}$.

This procedure can be extended to fits of molecular configurations.  In
this case the deviations from the mean configuration,
eq.~(\ref{meanconfig}), is mapped into a point in $\mathbb R^6$; the
resulting Gaussian fit will capture the correlation between the
translational and rotational degrees of freedom.

In closing this section, we mention an alternative way of fitting
quaternion orientational data with analytic functions, namely in terms
of spherical harmonics.  The normal (3-dimensional) spherical harmonics
can be generalized to 4 (and higher) dimensions
\cite{mueller66,mueller98} and the orthogonality relation allows the
coefficients of the harmonics to be computed simply.  The $\pm\q q$
symmetry merely results in the odd harmonics dropping out.  However in
typical molecular interactions, the relative orientation of the
molecules is tightly constrained which means that a large number of
spherical harmonics will be needed to represent the orientational
distribution.  For such applications, a representation in terms of
localized functions, such as Gaussians, is preferable.

\section{Random orientation}

In Monte Carlo simulations \cite{metropolis53} it is sometimes necessary
to select a molecule with a random and uniform position and orientation,
for example, when attempting to insert a molecule into a simulation box
during a grand canonical simulation \cite{adams75}.  Choosing a random
position is straightforward.  However, we need to be careful to select
the random orientation uniformly or else detailed balance will be
violated (when balancing insertions and deletions).  One possibility is
to choose a random turn $\v u$ in $\mathbb B^3$ and to convert this to a
quaternion.  However, it is much simpler to sample directly in
quaternion space.

Let us first establish the requirement for ``uniform'' sampling of
orientations.  Composing 3-dimensional rotations is carried out by the
multiplication of unit quaternions; but we know that $\q p \q q = \q
p\transp \cdot \Q Q(\q q)$, where $\Q Q(\q q)$, given in
eq.~(\ref{rightmult}), is orthonormal if $\q q$ is a unit quaternion.
Thus 3-dimensional rotations map into a rigid rotation of $\mathbb S^3$;
a uniform density on $\mathbb S^3$ is invariant to such rotations.  It
follows that the task of sampling a random orientation reduces to
picking a random unit quaternion uniformly on $\mathbb S^3$.

Marsaglia \cite{marsaglia72} provides one prescription: select $x_1$ and
$y_1$ uniformly in $(-1,1)$ until $s_1 = x_1^2 + y_1^2 < 1$; similarly,
select $x_2$ and $y_2$ uniformly in $(-1,1)$ until $s_2 = x_2^2 + y_2^2
< 1$; then
\[
\q q = [ x_1, y_1, x_2 \sqrt{(1-s_1)/s_2}, y_2 \sqrt{(1-s_1)/s_2} ]
\]
is uniformly distributed on $\mathbb S^3$.

A more transparent and symmetric method (which generalizes to sampling
points on $\mathbb S^n$ \cite[\S7.1]{muller56}) is to pick 4 normal
deviates $g_i$ for $i \in [0,4)$ and to set
\[
\q p = [ g_0, g_1, g_2, g_3], \quad
\q q = \q p/\abs{\q p}.
\]
Although this method is less efficient than Marsaglia's, the overall
impact in the context of a molecular simulation is probably tiny.  Both
of these methods return points uniformly over the whole of $\mathbb S^3$
rather that over just one hemisphere.  In most applications, this is of
no consequence.

Other representations of rotation yield more complex rules for obtaining
random orientations.  For example, with Euler angles, we would sample
uniformly the first and third angles and the cosine of the second angle.
If the orientation is given in axis-angle space, $\v s$, then the axis,
$\v s/\abs{\v s}$, should be chosen uniformly on $\mathbb S^2$, and the
rotation angle, $\abs{\v s}$, should be sampled from $[0,\pi]$ with
probability $(2/\pi)\sin^2(\abs{\v s}/2)$.  Of course, this simplifies
when $\v s$ is transformed to $\v u$ space, eq.~(\ref{lambert}), leading
to a uniform distribution in $\mathbb B^3$.

A related problem is selection of random rotational moves for use in a
Monte Carlo simulation \cite{metropolis53}.  This method requires that
detailed balance be satisfied, which, in the absence of torque bias,
means that the probability of selecting the new orientation is symmetric
under interchange of old and new orientations.  Because we are typically
interested in small changes in orientation, it is most convenient to
select the rotation in axis-angle space as $\exp([0, \v s])$ and to set
the new orientation
\[
\q q' = \exp([0, \v s])\q q,
\]
where $\v s$ is selected from an even distribution, $p(\v s) = p(-\v
s)$.  (This result follows because the Jacobian factor is even in $\v
s$.)  Usually, we wish the choice of rotation axis to be isotropic, and
in that case we have $p(\v s) = p(\abs{\v s})$.  Thus we might select
$\v s$ uniformly in a sphere of radius $\Delta$.  Rao {\em et al.}\
\cite{rao79} select $\abs{\v s}$ uniformly in $[0, \Delta]$ (which
results in a distribution which is singular at the origin in $\v s$
space).  An attractive choice of distribution is a 3-dimensional
Gaussian
\[
p(\v s) = \frac{\exp(-\frac12 \abs{\v s}^2/\Delta^2)}
               {(2\pi)^{3/2}\Delta^3}.
\]
Not only is this simple to sample from, but it allows torque bias to be
included in a simple manner.  Torque bias is implemented \cite{rao79} by
multiplying the {\em a priori} probability of selecting a move by
$\exp(\lambda\beta\, \v t\transp \cdot \v s)$, where $\beta$ is the
inverse temperature, $\lambda$ is a constant (usually taken to be
$\frac12$), and $\v t$ is the torque on the molecule.  If the ``starting''
distribution is a Gaussian then the torque-bias factor merely shifts the
Gaussian to give
\[
p(\v s) = \frac{\exp(-\frac12 
        \abs{\v s - \lambda\beta\Delta^2 \v t}^2/\Delta^2)}
               {(2\pi)^{3/2}\Delta^3}.
\]
This offers two simplifications over the original procedure
\cite{rao79}: (a)~it is trivial to sample from a shifted Gaussian; and
(b)~the acceptance probability, which involves the ratio of the forward
and reverse {\em a priori} probabilities, is also easy to compute and,
in particular, it does not require the evaluation of a normalization
factor for the distribution.  Similar considerations obviously apply to
the application of force bias for translational moves, as has been
discussed by Rossky {\em et al.}~\cite{rossky78}.  Indeed, in the case
of moving molecules, we would naturally perform a combined translational
and orientational move applying both force- and torque-bias
simultaneously.  There are often strong gradients in the forces in
molecular simulations and a direct application of force bias in this
case can lead to poor sampling because certain transitions are
effectively disallowed.  In such cases, it is prudent to limit the
effect of the bias by limiting the shift in the Gaussian, if necessary,
to ensure that there a finite probability (at least 5--10\%, say) of the
sampled move being in the opposite direction to the force.  This ensures
that the molecule can effectively explore configuration space because
small steps are always permitted and it provides a simpler ``safety''
mechanism than the distance scaling of $\lambda$ proposed by Mezei
\cite{mezei91}.

Finally, some care needs to be taken to treat the possibility of the
orientation ``wrapping'' around.  Suppose the sampled $\v s$ has
$\abs{\v s} > \pi$, then the resulting orientation is identical to the
wrapped one, $\v s - 2\pi \v s/\abs{\v s}$.  To ensure that detailed
balance is maintained, the acceptance probability should use the {\em a
priori} probability for the reverse move $-\v s$ (rather than the
negative of the wrapped move).  A simple expedient for avoiding this
problem is simply to reject any move with $\abs{\v s} > \pi$.

\section{Grids for orientation}

In many contexts, it is important to be able to represent the
independent variables for a problem on a grid.  It is therefore useful
to be able to map orientations onto a grid.  Possible applications are
binning molecular data, implementing cavity bias in orientation
\cite{mezei80}, performing systematic searching of orientations (where
the goal is to provide more regular coverage of orientation space than
is achieved by random sampling), and performing integrals over
orientation by numerical quadrature \cite{eden98}.
Our goal is to provide a simple
rule for covering orientation space with a grid while ensuring that the
grid elements are approximately of equal volumes and are not unduly
distorted.  Here again, representing the orientation as a quaternion
provides a reasonable solution.

Recall that unit quaternions lie on a hypersphere $\mathbb S^3$.
Positions on $\mathbb S^3$ can be determined by 3 angle-like variables.
However these are a poor basis for a grid because of singularities in
the resulting coordinate system.  Instead let imagine surrounding
$\mathbb S^3$ by a tesseract (the 4-dimensional analogue of the cube) of
edge length 2.  This consists of 8 cells which are $2 \times 2 \times 2$
cubes tangent to $\mathbb S^3$.  An exemplary cell is given by $\q p$
with $p_0 = 1$, $\abs{p_{i\ne0}} \le 1$.  We need only consider half of
the cells of the tesseract because of the identification of $\pm \q p$.
Thus we choose to consider the four cells for which one of the
components of $\q p$ is $+1$.

This then forms the basis for a cubical grid for orientation space.
This is attractive because cubical grids are simple to index into; they
are easy to refine; they have an metric factor which is easy to compute;
etc.  The overall ``wastefulness'' of this grids relative to a cubic
grid within a domain of $\mathbb R^3$ is given by the ratio of the
volume of four cells of the tesseract ($4\times 2^3$)
to the area (really a volume) of a
hemisphere of $\mathbb S^3$, i.e., $32/\pi^2 \approx 3.24$.
This might seem rather profligate.  However, if we managed to arrange
the grid around the $\mathbb S^3$ without any wastage, the grid edge
would be reduced by a factor of only $\sqrt[3]{3.24} \approx 1.48$.

Let us divide each of the cells of the tesseract into $M^3$ grid cubes
(of side $2/M$).  These cubes can then be projected to $\mathbb S^3$ by
scaling $\q p$ to a unit quaternion.  This operation scales the volume
of each of the grid cubes by $\abs{\q p}^{-4}$---a factor of $\abs{\q
p}^{-3}$ is due to scaling a volume element linearly by $\abs{\q
p}^{-1}$ and the last factor of $\abs{\q p}^{-1}$ arises from the
distortion of the cube during this operation.  The maximum scaling
occurs at the corners of the tesseract, e.g., $\q p = [1,1,1,1]$, where
$\abs{\q p}=2$, so that range of volumes for the grid elements is 16
with the maximum distortion being a factor of 2.  Mapping between an
arbitrary orientation $\q q$ and a point in the grid is then achieved as
follows.  We identify the component $q_l$ of $\q q$ which is largest in
absolute value and set $\q p = \q q/q_l$, giving $p_l = 1$ and $p_{i\ne
l} \in [-1,1]$.  The grid then consists of $4\times M\times M\times M$
elements.  The resolution of the grid, given by the maximum change in
orientations between neighboring grid cells, is
approximately $4/M$.  (We need to multiply the grid cube edge by 2 to
obtain the equivalent rotation angle, because, from eq.~(\ref{rot0}), we
have $\q q = [1, \v v \theta/2]$ for $\theta$ small.)

When the application is quadrature, it is natural to evaluate the
function and to compute the metric factor $\abs{\q p}^{-4}$ at the
centers of the grid cubes.  For binning, we assign the samples to the
grid cube in the obvious way and again use the grid center to compute
the metric factor to obtain a sample density.

The cubical grid defined above is suitable for quadrature and
searching where the cost of function evaluations is small.  Sometimes,
however, the cost of function evaluations is so high that it is
desirable to find an ``optimal'' set of grid points.  For integrations
over $\mathbb S^2$, this is a well-studied problem
\cite{mclaren63} and various integration grid have been given that
ensure accuracy to high order \cite{lebedev99}.  For $\mathbb S^3$,
various spherical $t$-designs are known \cite{delsarte77,sloane03}.  A
$t$-design is a set of points on the sphere such that the average of a
polynomial of degree $t$ over the sphere is given by averaging the
value of polynomial at those points.  Unfortunately $t$-designs for
$\mathbb S^3$ are only known for $t$ up to 11 with the 11-design
corresponding to 60 orientations.

In order to provide a denser coverage of the sphere we propose the
following strategy: Consider $N$ sample orientations, corresponding to
$2N$ points on $\mathbb S^3$.  Define a ``covering radius'', $\alpha$,
as the maximum rotation needed to align an arbitrary orientation with
one of the sample orientations.  The ``coverage'', $c$, is defined by
the ratio of the area of $2N$ spherical caps of rotational extent
$\alpha$ to the total area of $\mathbb S^3$, i.e.,
\begin{equation}\label{coverage}
c = \frac{N(\alpha-\sin\alpha)}\pi
\end{equation}
[compare with eq.~(\ref{lambert}b)].  For a given $N$, the optimal
configuration of sample orientations is obtained by minimizing
$\alpha$---this gives the ``thinnest'' coverage, $c$.  Finally, we
weight each sample point according to the fraction of orientational
space which is closest to it (i.e., in proportion to the volumes of the
Voronoi cells); and we set a secondary goal of minimizing the variation
in the weights.  We expect the resulting sample points and weights to
provide robust and accurate estimates of orientational
integrals---particularly of experimentally or numerically determined
functions which are bounded but which may not have bounded derivatives.
The sample points are also suitable for searching orientation space
optimally.  Finding such optimal sets of points is difficult in
practice.  So, here, we propose some sets based on the regular
4-dimensional polytopes \nocite{coxeter63}\cite{coxeter63,bourke03},
with the results summarized in table~\ref{covering}.
\begin{table}\ifeprint\else\renewcommand{\baselinestretch}{1.1}\fi
\caption{Coverings of orientation space.  $N$ is the number of
orientations; $\alpha$ is the covering radius (expressed as a
rotation), and $c$ is the coverage, eq.~(\ref{coverage}).}\label{covering}
\begin{tabular}{\ifeprint\else@{\extracolsep{2em}}\fi cccl}
\hline\hline
$N$ & $\alpha$ & $c$ & notes\\\hline
   24&  $62.80^\circ$\footnote
{$\cos^{-1} \bigl( \frac14 (2\sqrt2 - 1) \bigr)$.}
                     &  1.579& vertices of 2 24-cells,
                        a 7-design\footnote{See \cite{sloane03}.}
\\
   60&  $44.48^\circ$\footnote
{$\cos^{-1} \bigl( \frac18 (3\sqrt5 - 1) \bigr)$.}
                     &  1.445& vertices of 600-cell,
                        an 11-design\footnotemark[2]\\
  360&  $27.78^\circ$&  2.152& vertices and cells of 600-cell\footnote
{60 vertices with weight 1.32870 and 300 cell centers with weight
0.93426.}\\\hline
   50&  $69.66^\circ$&  4.426& ZCW3\_50\footnote{Euler angles for
   orientations taken from \cite{eden98}.}\\
  538&  $32.53^\circ$&  5.142& ZCW3\_538\footnotemark[5]\\
 6044&  $18.10^\circ$& 10.051& ZCW3\_6044\footnotemark[5]\\\hline
$\infty$& $5.499/\sqrt[3]N$& 8.821\footnote
{$16\sqrt3/\pi$.}& cubic lattice in tesseract\footnote
{Ratio of maximum to minimum weights is $16$.}\\
$\infty$& $4.472/\sqrt[3]N$& 4.745\footnote
{$20\sqrt5/(3\pi)$.}& body-centered cubic lattice in
tesseract\footnotemark[7]\\
$\infty$& $4.092/\sqrt[3]N$& 3.635\footnote
{$224\sqrt3/[(17+12\sqrt2)\pi]$.}& cubic lattice in 48-cell\footnote
{Ignoring boundary effects, ratio of maximum to minimum weights is
$64/(17+12\sqrt2) = 1.884$.}\\
$\infty$& $3.328/\sqrt[3]N$& 1.956\footnote
{$280\sqrt5/[3(17+12\sqrt2)\pi]$.}& body-centered cubic lattice in
48-cell\footnotemark[10]\\
   648 & $20.83^\circ$ & 1.641 & c48u27\footnote
{Body-centered cubic lattice in a 48-cell with lattice spacing $\delta$; see
\cite{karney06}.} ($\delta=0.33582$)\\
  7416 & $10.07^\circ$ & 2.133 & c48u309\footnotemark[12] ($\delta=0.15846$)\\
 70728 & $ 4.71^\circ$ & 2.078 & c48u2947\footnotemark[12] ($\delta=0.07359$)\\
$\infty$& $3.022/\sqrt[3]N$& 1.464\footnote
{$5\sqrt5\pi/24$.}& uniform body-centered cubic lattice\footnote
{Conjectured thinnest covering for $N\rightarrow\infty$, based on
optimal covering of $\mathbb R^3$
\cite{bambah54}.}\\\hline\hline
\end{tabular}
\end{table}

The 24-orientation set is obtained by placing two 24-cells (or
icositetrachora) in their mutually dual configurations to give the set
\begin{subequations}\label{cubesym}
\begin{eqnarray}
\mbox{8 permutations of}&[\pm 1, 0, 0, 0],\\
\mbox{16 permutations of}&
\bigl[\pm \frac12,\pm \frac12,\pm \frac12,\pm \frac12\bigr],\\
\mbox{24 permutations of}&
\bigl[\pm \frac1{\sqrt2},\pm \frac1{\sqrt2}, 0, 0\bigr].
\end{eqnarray}
\end{subequations}
(Each orientation is counted twice here because of the identification of
$\pm\q q$.)  The corresponding Voronoi tessellation is a truncated-cubic
tetracontaoctachoron (or 48-cell) which consists of 48 regular truncated
cubes \cite{olshevsky04}.  The set of orientations, eq.~(\ref{cubesym}),
is the direct symmetry group for the cube.

The vertices of the 600-cell (or hexacosichoron) \cite{bourke03} are
given by eqs.~(\ref{cubesym}a) and (\ref{cubesym}b) together with
\begin{eqnarray*}
\mbox{96 even permutations of}&
\bigl[\pm \frac{\sqrt5 + 1}4,\pm \frac{\sqrt5 - 1}4, \pm\frac12, 0\bigr].
\end{eqnarray*}
In this case, the Voronoi tessellation is the dual of the 600-cell,
namely the 120-cell (or hecatonicosachoron).  Because the Voronoi cells
are dodecahedra which are nearly spherical, the resulting 60
orientations gives a particularly thin covering of orientation space.
A good covering is also provided by adding the centers of the
tetrahedral cells of the 600-cell.

For comparison, we list in table~\ref{covering} the data for some of the ZCW3
orientation sets used by Ed\'en and Levitt \cite{eden98}.  These are
obtained by the taking sets of points appropriate for integrating of a
periodic unit cube
\nocite{zaremba66}\nocite{conroy67}\cite{zaremba66,conroy67,cheng73}
and mapping this set to the space of 3 Euler angles.
There are two potential problems with this approach: (1) even though the
metric of orientation space is treated properly, the mapping from Euler
angles to orientation space is not distance-preserving and we expect
this to degrade the properties of a mesh; and (2) because one of the
Euler angles is not periodic, functions in orientation space do not obey
the constraints assumed in constructing the sets of sample points.
(More complete data for the ZCW3 sets is available in \cite{karney06}.)

Finally, table~\ref{covering} provides various strategies for constructing
an arbitrarily fine grid.  We start with gridding the tesseract on which
we easily impose a cubical grid (see above).  However, the optimal
sphere covering of $\mathbb R^3$ is body-centered cubic \cite{bambah54},
and such a grid results in a thinner covering.  Still better coverings
can be found by starting with the 48-cell which has a typical cell (a
truncated cube),
\begin{equation}\label{trunccube}
p_0 = 1,\quad
\abs{p_{i\ne0}} \le \sqrt2-1,
\quad\abs{p_1} + \abs{p_2} + \abs{p_3} \le 1.
\end{equation}
The other cells are obtained by multiplying $\q p$ by the members of
eq.~(\ref{cubesym}).  A cubic or body-centered cubic grid can easily
be placed within each cell.
For example, a body-centered lattice can be obtained with
\[
p_0 = 1, \quad \v p = [k, l, m]\delta/2,
\]
subject to the constraint eq.~(\ref{trunccube}), where $k$, $l$, and $m$
are either all even integers or all odd integers.  Table~\ref{covering}
gives three examples of such grids.
The disadvantage of grids in 48-cells is that
care must be taken to treat the faces of the cell correctly.  The
triangular faces of the truncated cubes slice cut through the grid
cells at an angle and the octagonal faces fit together with a
$45^\circ$ twist.  It is therefore necessary to resort to numerical
methods to determine the volume of the Voronoi cells near the faces.
The resulting data for the weights and examples of other body-centered
cubic grids in the 48-cell with $\alpha \ge 0.65^\circ$ are given in
\cite{karney06}.

One special searching problem is determining the volume of the
smallest rectangular box (whose edges are parallel to the coordinate
axes) into which a given molecule fits.  This problem arises in the
study of a single protein bathed in a solvent.  In order to eliminate
boundary effects, it is possible to construct a periodic system and,
for efficiency, we wish the volume of the periodic cell to be minimum.
We can solve this problem by systematically sampling over all
orientations using our grid.  However, because of the symmetries of a
cube, eq.~(\ref{cubesym}), there are 24 equivalent orientations which
minimize the volume and we can restrict the search to $1/24$ of
orientation space by searching only in eq.~(\ref{trunccube}).  We
should point out that for the purposes of mimicking a single solute
molecule in a solvent with a periodic system, the ``best''
computational box is not given by fitting a single image of the solute
into a box but rather by the more challenging problem of optimally
fitting the solute molecules into its neighboring images
\cite{bekker04}.

The emphasis in the section is on covering all orientation space with a
grid.  In many molecular modeling applications, the orientation may be
quite restricted, e.g., when considering the orientation of a ligand in
a protein binding pocket, and we may elect to restrict the integration
(or search) to a set of orientations which differ from the mean rotation
by at most $\Theta$.  If we express the deviation from the mean as a
turn vector, eq.~(\ref{lambert}), integrations may be carried out in
(three-dimensional) turn space with the range of integration restricted
to the ball $\abs{\v u} \le \sqrt[3]{(\Theta-\sin\Theta)/\pi}$.  Because
the mapping to turn space is volume preserving, the integrals are exact.
In addition, provided that $\Theta \lesssim \pi/2$, the mapping to turn
space entails little distortion ($\lesssim 12\%$) and standard numerical
methods for integrating in a ball $\mathbb B^3$ can be used.

\section{Discussion}

Quaternions are an ideal ``fiducial'' representation \cite{kahan98} of
orientation in a molecular simulation.  They provide an economical format
for program input and output and as the internal representation of
orientation.  There is little redundancy in the representation---there
is just the normalization constraint on its four elements and this is
easily tested and corrected.  At a given numerical precision,
quaternions cover orientation space uniformly.  Most operations
involving orientation can be carried out directly and efficiently with
quaternions and they can be converted to other representations as
needed.  The basic operation of composing rotations is most cheaply
performed with quaternions.  On the other hand, if we need to rotate a
large molecule it is quicker to convert the quaternion to a rotation
matrix, eq.~(\ref{mat3}), and to perform matrix-vector multiplication than
to apply eq.~(\ref{rot}) directly.

In comparison, other representation suffer serious drawbacks.
Rotations cannot be easily composed when expressed as Euler angles.  
Picking a random orientation is more awkward when rotation matrices
are used.  In neither of these representations is it easy to interpolate
between two orientations or to compute the mean orientation.

Although quaternions may be unfamiliar to some readers, we only needed
to use quaternion algebra in the rule for composing rotations and in
deriving the least squares fit.  In carrying out the other tasks, we
just used the fact that rotations are represented by opposite points on
$\mathbb S^3$ and this provides a ``natural'' metric for rotations.  In
working with $\mathbb S^3$, we are able to carry over geometrical
concepts from $\mathbb S^2$ or use straightforward extensions from
Euclidean space, $\mathbb R^3$, to $\mathbb R^4$.

A curious and non-obvious property of rotations which is evident from
their representation on $\mathbb S^3$, with $\pm\q q$ identified, is
that rotations do not form a simply connected group.  Thus, if we rotate
an object by $360^\circ$ it returns to its original orientation but with
the sign of $\q q$ changed.  This means that we cannot continuously
deform the path that the object took to reduce it to a point.  However,
we {\em can} do this if we rotate the object by $720^\circ$.  This
property of rotations is an immediate consequence of their
representation as a pair of points $\pm\q q$ on $\mathbb S^3$ and good
visual illustrations of this property are provided by the Dirac belt
trick \cite{egan00} and the Phillipine wine dance \cite{francis93}.

\section*{Acknowledgment}

This work was supported by the
\hrefx{https://mrmc-www.army.mil}{U.S. Army Medical Re-}{search and Materiel
Command} under Contract No.\ DAMD17-03-C-0082.  The views, opinions, and
findings contained in this report are those of the author and should not
be construed as an official Department of the Army position, policy, or
decision.  No animal testing was conducted and no recombinant DNA was
used.

\bibliography{free}
\end{document}